\documentclass[aps,pre,twocolumn,showpacs]{revtex4-1}
\usepackage{epsfig}
\usepackage{bm}

\newcommand{\be}{\begin{eqnarray}}
\newcommand{\ee}{\end{eqnarray}}
\newcommand{\ba}{\begin{array}}
\newcommand{\ea}{\end{array}}
\newcommand{\bmat}{\left(\begin{array}}
\newcommand{\emat}{\end{array}\right)}
\newcommand{\no}{\nonumber}
\newcommand{\Tr}{\mbox{Tr}\,}
\newcommand{\diff}{\mathrm d}
\newcommand{\e}{\mathrm e}

\begin{document}
\title{Optimization of Markov process violates detailed balance condition}
\author{Kazutaka Takahashi}
\affiliation{Department of Physics, Tokyo Institute of Technology, 
Tokyo 152-8551, Japan}
\author{Masayuki Ohzeki}
\affiliation{Department of Systems Science, Graduate School of Informatics,
Kyoto University, Kyoto 606-8501, Japan}

\date{\today}

\begin{abstract}
We consider the optimization of Markovian dynamics to pursue 
the fastest convergence to the stationary state.
The brachistochrone method is applied to the continuous-time master equation 
for finite-size systems.
The principle of least action leads to a brachistochrone equation 
for the transition-rate matrix.
Three-state systems are explicitly analyzed, and we find that 
the solution violates the detailed balance condition.
The properties of the solution are studied in detail to observe 
the optimality of the solution.
We also discuss the counterdiabatic driving for the Markovian dynamics.
The transition-rate matrix is then divided into two parts, and 
the state is given by an eigenstate of the first part.
The second part violates the detailed balance condition and 
plays the role of a counterdiabatic term.
\end{abstract}
\maketitle

\section{Introduction}

The Markov-chain Monte Carlo method is widely used in various fields of research 
to realize a desired distribution by simulating the stochastic dynamics.
Frequently used algorithms such as Metropolis and heat-bath algorithms
are under a well-known limitation 
called the detailed balance condition (DBC)~\cite{MRRTT, Hastings}.
This is not a necessary condition but rather a sufficient one 
for convergence to the stationary distribution.
Recent studies revealed that violation of the DBC can accelerate relaxation.
Several algorithms were proposed in the literature to show 
the speed-up of convergence to the stationary state 
and the reduction of the correlation time of the sampling~\cite{ST, TCV, FW, SH}.
Ichiki and Ohzeki mathematically showed that the violation gives shorter 
relaxation times~\cite{IO1}.
They provided insight into the acceleration of the relaxation to 
the steady state from the viewpoint of rare-event sampling~\cite{IO2} 
and developed a simple method to implement the violation 
in Langevin dynamics~\cite{OI,OI2}.

The preceding studies, as listed above, focused on the violation of the DBC.
Although the violation actually leads to the speed-up of the convergence, 
we still have considerable freedom in choosing the algorithm.
No general principles to determine the optimality of algorithms are known, 
which prevents us from developing efficient algorithms.
Several studies on optimization were performed within the DBC~\cite{Liu, PRHH}.
In addition, the minimization of the Kullback--Leibler divergence 
between the original dynamics under the DBC and 
an alternative stochastic pathway results in a nontrivial solution without the DBC 
but with the common observables in the stationary state~\cite{IO2,SO}.
In the present study, we formulate a general method to generate 
the optimal stochastic rule by developing an optimization method 
in quantum mechanics.

The Markov-chain Monte Carlo method is formulated using the master equation.
It is a first-order differential equation in time and 
is formally understood as the imaginary-time Schr\"odinger equation.
The transition-rate matrix corresponds to the quantum Hamiltonian.
We exploit this similarity to develop a formal method 
to optimize the stochastic dynamics.
In quantum systems, the brachistochrone method has been studied 
to determine the optimal Hamiltonian under specified constraints~\cite{CHKO, CHKO2}.
We may naively expect that this idea can be directly applied to 
the optimization of the master equation.
The optimal solution yields the stochastic dynamics with 
the fastest convergence to the stationary state, which results in 
the best algorithm in the Markov-chain Monte Carlo method 
in terms of convergence speed.

Furthermore, the quantum brachistochrone method leads to 
a fascinating connection to a different type of optimization  
called ``shortcuts to adiabaticity''~\cite{DR1, DR2, Berry, CRSCGM, STA}.
Under some condition, the optimized quantum state obtained from 
the brachistochrone equation (BE) follows an adiabatic passage, 
which is defined by the instantaneous eigenstates of 
a time-dependent Hamiltonian~\cite{Takahashi}.
Although we usually treat a time-independent transition-rate matrix 
in the master equation, studying counterdiabatic driving 
in the present system can provide insight to obtain the optimal algorithm.

The aim of the present work is as follows. 
First, we formulate a brachistochrone method to optimize 
the transition-rate matrix in the master equation.
Second, we investigate whether the DBC is actually violated 
in the optimized solution.
Third, we study our optimization method 
from the viewpoint of shortcuts to adiabaticity.

The rest of the paper is organized as follows.
In Sec.~\ref{sec:meq}, the master equation is defined 
and is deformed for the convenience of our analysis.
Section~\ref{sec:be} is the main part of the present work, where 
the brachistochrone method for the master equation is formulated.
The general form of the BE is derived, and we discuss its general properties.
Three-state systems are studied in detail in Sec.~\ref{sec:three}.
In Sec.~\ref{sec:sta}, we extend our method to time-dependent systems to observe 
the connection to shortcuts to adiabaticity.
The summary is presented in the last section~\ref{sec:sum}.

\section{Master equation}
\label{sec:meq}

The Markov-chain Monte Carlo method is described by the master equation.
The continuous-time master equation for discrete $N$ states is suitable 
for our purpose and is written as 
\be
 \frac{\diff}{\diff t}p_i(t) = \sum_{j=1}^N q_{ij}p_j(t), \label{meq0}
\ee
where $i=1,2,\dots,N$.
$p(t)=(p_1(t), p_2(t), \dots, p_N(t))$ denotes a probability distribution 
at time $t$ satisfying
\be
 \sum_{i=1}^N p_i(t)=1. \label{norm0}
\ee
The coefficient $q_{ij}$ with $i\ne j$ is positive and denotes 
an incoming transition probability rate from the state $j$ to $i$.
The diagonal element $q_{ii}$ is negative, and the magnitude represents 
the outgoing probability rate of state $i$. 
It is determined so that the following relation holds: 
\be
 \sum_{j=1}^N q_{ji}=0. \label{qii}
\ee
By taking the sum over $i$ in Eq.~(\ref{meq0}), we see that this is a condition 
for the conservation of normalization (\ref{norm0}).

At $t\to\infty$, we assume that the probability distribution becomes 
a stationary one as  
\be
 p_i(t)\to \pi_i.
\ee
Then, the balance condition requires 
\be
 \sum_{j=1}^N q_{ij}\pi_j =0. \label{bc0}
\ee
The transition-rate matrix $q$ is determined so that Eqs.~(\ref{qii}) 
and (\ref{bc0}) are satisfied.
In the traditional approach, we use the DBC 
\be
 && q_{ij}\pi_j = q_{ji}\pi_i \label{dbc0}
\ee
instead of the balance condition (\ref{bc0}).
This is a sufficient condition for convergence to the stationary 
distribution and is not necessarily satisfied.
In the present study, we consider the optimization of the master equation 
without imposing the DBC.
For convergence to the stationary distribution, 
we need an additional condition of irreducibility.
Irreducibility can be checked by observing the obtained solution, and 
we do not treat the condition explicitly.

For the convenience of notation, we change the variables as 
\be
 && P_i(t) = \frac{1}{\sqrt{\pi_i}}p_i(t), \\
 && W_{ij} = \frac{1}{\sqrt{\pi_i}}q_{ij}\sqrt{\pi_j}.
\ee
Then, the form of the master equation is unchanged:  
\be
 \frac{\diff}{\diff t}P_i(t)=\sum_{j=1}^N W_{ij}P_j(t). \label{meq}
\ee
The normalization condition for $P_i(t)$ is given by
\be
 \sum_{i=1}^N\sqrt{\pi_i}P_i(t)=1. \label{norm}
\ee
By definition, the transition rate $W_{ij}$ satisfies 
\be
 \sum_{i=1}^N\sqrt{\pi_i}W_{ij}=0. \label{sbd}
\ee
This condition is equivalent to Eq.~(\ref{qii}).
We note that the diagonal elements of $W$ are negative and 
the off-diagonal elements are positive.

The stationary distribution is given by 
\be
 P_i(t)\to\sqrt{\pi_i}.
\ee
The balance condition (\ref{bc0}) and DBC (\ref{dbc0}) are rewritten, 
respectively, as 
\be
 && \sum_{j=1}^N W_{ij}\sqrt{\pi_j}=0, \label{bc} \\
 && W_{ij}=W_{ji}.
\ee
The advantage of representation (\ref{meq}) is that 
the DBC is represented simply by the symmetry of the matrix $W$.
In the following, we consider the formulation based on this representation.

Since the transition-rate matrix $W$ is asymmetric in general, 
the eigenvalue equations are written by using the right and left eigenstates. 
\be
 && W|R_i\rangle = \Lambda_i|R_i\rangle, \\
 && \langle L_i|W = \langle L_i|\Lambda_i.
\ee
These vectors are orthonormalized with each other: 
\be
 \langle L_i|R_j\rangle = \delta_{ij}.
\ee
We also know that 
$|\pi\rangle=(\sqrt{\pi}_1,\sqrt{\pi}_2,\ldots,\sqrt{\pi}_N)^{\rm T}$ and 
$\langle\pi|=(\sqrt{\pi}_1,\sqrt{\pi}_2,\ldots,\sqrt{\pi}_N)$ 
are the right and left eigenstates with zero eigenvalue, respectively.
They satisfy the normalization $\langle \pi|\pi\rangle = 1$ and 
are orthogonal to the other eigenstates.
The spectral decomposition of $W$ is written as 
\be
 W = \sum_{i=1}^{N-1}\Lambda_i|R_i\rangle\langle L_i|
 +0\cdot|\pi\rangle\langle\pi|.
\ee
The formal solution of the probability distribution
$|P(t)\rangle=(P_1(t),\ldots,P_N(t))^{\rm T}$ is given by 
\be
 |P(t)\rangle &=& \e^{Wt}|P(0)\rangle \no\\
 &=& \sum_{i=1}^{N-1}\e^{\Lambda_it}
 |R_i\rangle\langle L_i|P(0)\rangle
 +|\pi\rangle, \label{pt}
\ee
where we use the property $\langle \pi|P(t)\rangle=1$.
In order for the system to approach the stationary distribution $\pi$, 
we require the condition ${\rm Re}\,\Lambda_i <0$.
This property is mathematically described by the Perron--Frobenius theorem.

Representation (\ref{pt}) shows that 
the relaxation time is given by ${\rm Max}\,(-1/{\rm Re}\,\Lambda_i)$.
Smaller values of the relaxation time accelerate the relaxation 
to the stationary distribution. 

\section{Brachistochrone equation}
\label{sec:be}

\subsection{Kullback--Leibler divergence}

The main idea of the present work is to use the similarity between 
the master equation and the Schr\"odinger equation. 
We naively expect that the formulation used in quantum mechanics can be 
directly applied to the Markovian dynamics.
The brachistochrone method requires a quantity to be optimized 
using the variational principle.
Here, we optimize the duration between the initial and final states.
This means that our aim is to derive the optimal solution with the fastest convergence 
to the stationary distribution.
The time is represented as the distance divided by the velocity and 
is written in an integral form.
In quantum mechanics, to define the distance between two states in Hilbert space, 
we exploit the Fubini--Study distance as a natural measure~\cite{CHKO}.
In the present problem, we need the distance in space of the probability 
distribution, which is not the same as the distance in Hilbert space.
We introduce the Kullback--Leibler divergence, 
\be
 D(p|q)=\sum_{i=1}^N p_i\ln\left(\frac{p_i}{q_i}\right),
\ee 
to represent the distance between the probability 
distributions $p=(p_1,p_2,\dots, p_N)$ and $q=(q_1,q_2,\dots,q_N)$. 
This is a well-known quantity in probability theory and 
is exploited as a natural measure for our study.

We consider the distance between $p$ and $q=p+\delta p$.
Since they represent probability distributions, the constraint 
\be
 \sum_{i=1}^N \delta p_i =0 
\ee
is imposed on $\delta p_i$.
For small $\delta p_i$, we obtain 
\be
 D = -\sum_{i=1}^N p_i\ln\left(1+\frac{\delta p_i}{p_i}\right) 
 \sim \sum_{i=1}^N\frac{(\delta p_i)^2}{2p_i}.
\ee
Thus, we can define the infinitesimal distance $\diff s$ as 
\be
 \diff s^2 = \sum_{i=1}^N\frac{(\delta p_i)^2}{2p_i}
 = \sum_{i=1}^N\sqrt{\pi_i}\frac{(\delta P_i)^2}{2P_i}.
\ee

Now that we have defined the distance, the velocity is naturally defined 
by combining the definition of the distance with the time-evolution law.
The deviation of the probability distribution after the small time 
evolution $t\to t+\delta t$ is obtained from the master equation as 
\be
 \delta P_i = \sum_{j=1}^N W_{ij}P_j\delta t.
\ee
The Kullback--Leibler divergence is written as 
\be
 D \sim v^2\delta t^2 
\ee
to define the velocity 
\be
 v = \sqrt{\sum_{i=1}^N\frac{\sqrt{\pi_i}}{2P_i}
 \Biggl(\sum_{j=1}^N W_{ij}P_j\Biggr)^2}.
\ee

By combining the expressions of the distance and velocity, 
we obtain the duration in an integral form as 
\be
 T = \int\frac{\diff s}{v} 
 = \int \diff t\,\sqrt{
 \frac{\sum_{i=1}^N\sqrt{\pi_i}\frac{\dot{P}_i^2}{P_i}}
 {\sum_{i=1}^N\sqrt{\pi_i}\frac{(\sum_{j=1}^NW_{ij}P_j)^2}{P_i}}}.
 \label{T}
\ee
We note that the master equation is not imposed in this expression.
When we impose the master equation, the integrand tends to unity, and
we obtain a trivial integral.
To determine the optimal solution, we perform the variation in a larger space.
The master equation is imposed as a constraint.

In the case of the quantum brachistochrone method, 
the form of the BE is shown to be insensitive 
to the measure~\cite{CHKO2}.
It is not obvious whether the same property holds
in the present formulation.
Therefore, we adopt the above measure for optimization.

\subsection{Brachistochrone equation}

In the brachistochrone method, the Euler--Lagrange equation of motion is 
derived by considering the variation with respect to dynamical variables.
In the present case, we consider the variations with respect to 
the probability distribution $P(t)$ and the transition-rate matrix $W$ 
under some constraints.
The constraints to be imposed are as follows: 
(i) master equation (\ref{meq});
(ii) normalization (\ref{norm}); 
(iii) conservation of normalization (\ref{sbd}); 
(iv) balance condition (\ref{bc}); and
(v) other constraints for $W$ represented as 
\be
 f_a(W)=0 \quad (a=1,2,\ldots). \label{fa}
\ee
By introducing multipliers, 
we define the action to be minimized as $S=T+S_{\rm c}$, where 
\be
 S_{\rm c} &=& \int \diff t\,\sum_{i=1}^N\lambda_i^{({\rm i})}(t)\Biggl(
 \frac{\diff}{\diff t}P_i(t)-\sum_{j=1}^NW_{ij}P_j(t)\Biggr) \no\\
 && +\int \diff t\,\lambda^{({\rm ii})}(t)\left(1-\sum_{i=1}^N
 \sqrt{\pi_i}P_{i}(t)\right) \no\\
 && +\sum_{i,j=1}^N\left(\sqrt{\pi_i}W_{ij}\lambda_j^{({\rm iii})}
 +\lambda_i^{({\rm iv})}W_{ij}\sqrt{\pi_j}\right) \no\\
 && +\sum_a\lambda_a^{({\rm v})} f_a(W). 
\ee
We note that the multipliers $\lambda_i^{({\rm i})}(t)$ and $\lambda^{({\rm ii})}(t)$ 
are real functions of $t$ and the other multipliers,  
$\lambda_j^{({\rm iii})}$, $\lambda_j^{({\rm iv})}$, and 
$\lambda_a^{({\rm v})}$, are time-independent real constants.

We first consider the variation $P_i\to P_i+\delta P_i$.
The condition $\frac{\delta S}{\delta P_i(t)}=0$ yields 
\be
 && -\frac{\diff}{\diff  t}\left(
 \frac{\sqrt{\pi_i}\frac{\dot{P}_i}{P_i}}{\sum_{j=1}^N
 \sqrt{\pi_j}\frac{\dot{P}_j^2}{P_j}}
 \right)
 -\frac{\sum_{j=1}^N\sqrt{\pi_j}\frac{\dot{P}_j}{P_j}W_{ji}}
 {\sum_{k=1}^N\sqrt{\pi_k}\frac{\dot{P}_k^2}{P_k}}
 \no\\ 
 && -\dot{\lambda}_i^{({\rm i})}
 -\sum_{j=1}^N\lambda_j^{({\rm i})}W_{ji}-\lambda^{({\rm ii})}\sqrt{\pi_i} = 0.
 \label{dsdp}
\ee
To eliminate the multiplier constant $\lambda^{({\rm ii})}$, 
we multiply $\sqrt{\pi_i}$ and take the sum over $i$.
Thus, we obtain 
\be
 \lambda^{({\rm ii})}= -\frac{\diff}{\diff  t}\left(
 \frac{\sum_{i=1}^N\pi_i\frac{\dot{P}_i}{P_i}}{\sum_{j=1}^N
 \sqrt{\pi_j}\frac{\dot{P}_j^2}{P_j}}
 \right)
 -\sum_{i=1}^N\dot{\lambda}_i^{({\rm i})}\sqrt{\pi_i}.
\ee
Equation (\ref{dsdp}) is written in vector form as
\be
 (1-Q)|\dot{\lambda}^{({\rm i})}\rangle+W^{\rm T}|\lambda^{({\rm i})}\rangle=
 -(1-Q)|\dot{\delta}\rangle-W^{\rm T}|\delta\rangle, \no\\
 \label{phidot}
\ee
where $|\lambda^{({\rm i})}\rangle$ is the $N$-component vector, the elements of which are 
given by $\lambda_1^{({\rm i})}, \lambda_2^{({\rm i})},\dots, \lambda_N^{({\rm i})}$, 
respectively.
Here, $|\delta\rangle$ is defined in the same manner, and the $i$th element is given by 
\be
 \delta_i = \frac{\sqrt{\pi_i}\frac{\dot{P}_i}{P_i}}
 {\sum_{j=1}^N\sqrt{\pi_j}\frac{\dot{P}_j^2}{P_j}}.
 \label{deltai}
\ee
We also define the projection matrix
\be
 && Q = |\pi\rangle\langle\pi|.
\ee
Then, Eq.~(\ref{phidot}) is formally solved as 
\be
 && |\lambda^{({\rm i})}(t)\rangle+|\delta(t)\rangle
 = \e^{-W^{\rm T}t}\left(|\lambda^{({\rm i})}(0)\rangle
 +|\delta(0)\rangle\right).
 \label{phi}
\ee
We note that $(1-Q)W^{\rm T}=W^{\rm T}$.
By the multiplication of $W^{\rm T}$, a state is projected onto 
the space excluding $|\pi\rangle$.

Next, we consider the variation $W_{ij}\to W_{ij}+\delta W_{ij}$ and 
impose the condition $\partial S/\partial W_{ij}=0$.
We obtain
\be
 && -\int\diff t\,\frac{\sqrt{\pi_i}\frac{\dot{P}_i}{P_i}P_j}
 {\sum_{k=1}^N\sqrt{\pi_k}\frac{\dot{P}_k^2}{P_k}}
 -\int \diff t\,\lambda_i^{({\rm i})}P_j \no\\
 && +\sqrt{\pi_i}\lambda_j^{({\rm iii})}+\lambda_i^{({\rm iv})}\sqrt{\pi_j}
 +\sum_a\lambda_a^{({\rm v})}\frac{\partial f_a}{\partial W_{ij}} = 0.
\ee
$\lambda_j^{({\rm iii})}$ and $\lambda_i^{({\rm iv})}$ can be eliminated 
using the corresponding constraints, as was the case for $\lambda^{({\rm i})}$.
Then, we have 
\be
 && \int\diff t\left[
 \lambda_i^{({\rm i})}+\delta_i
 -\sqrt{\pi_i}\sum_{k=1}^N\sqrt{\pi_k}(\lambda_k^{({\rm i})}+\delta_k)
 \right](P_j-\sqrt{\pi_j})
 \no\\
 &=&
 \sum_a\lambda_a^{({\rm v})}\Biggl(
 \frac{\partial f_a}{\partial W_{ij}}
 -\sum_{k=1}^N\frac{\partial f_a}{\partial W_{ik}}\sqrt{\pi_k}\sqrt{\pi_j}
 \no\\
 &&
 -\sqrt{\pi_i}\sum_{k=1}^N\sqrt{\pi_k}\frac{\partial f_a}{\partial W_{kj}}
 +\sqrt{\pi_i}\sum_{k,l=1}^N\sqrt{\pi_k}\frac{\partial f_a}{\partial W_{kl}}
 \sqrt{\pi_l}\sqrt{\pi_j}
\Biggr). \no\\
\ee
By inserting the solution of $\lambda_i^{({\rm i})}$ into Eq.~(\ref{phi}) 
to this equation, we obtain the result in matrix form, 
\be
 && (1-Q)F(1-Q) \no\\
 &=& \int \diff t\,\left(|P(t)\rangle-|\pi\rangle\right)
 \left(\langle\lambda^{({\rm i})}(0)|+\langle\delta(0)|\right)\e^{-Wt}(1-Q),
 \no\\
\ee
where $F$ is the $N\times N$ matrix, the elements of which are given by 
\be
 && F_{ij}=\sum_a\lambda_a^{({\rm v})}\frac{\partial f_a}{\partial W_{ji}}.
 \label{F}
\ee
Using the formal representation of the master equation (\ref{pt}), 
we obtain the BE 
\be
 && F_\perp = \int_0^{T} \diff t\,\e^{Wt}G_\perp\e^{-Wt},
 \label{beq}
\ee
 where
\be
 && F_\perp = (1-Q)F(1-Q), \\
 && G_\perp =\bigl(|P(0)\rangle-|\pi\rangle\bigr)\langle\lambda_\perp|,
\ee
and $|\lambda_\perp\rangle$ is an arbitrary and constant vector satisfying 
\be
 Q|\lambda_\perp\rangle = 0.
\ee
We also have
\be
 && QF_\perp = F_\perp Q = 0, \\
 && QG_\perp = G_\perp Q = 0.
\ee
The BE (\ref{beq}) is the main result of the present work.

\subsection{General properties of the brachistochrone equation}

The constraints (\ref{fa}) determine the form of the matrix $F$, and 
the BE (\ref{beq}) is solved under given initial and final states with 
constraints (i)--(v), in principle.
It should be noted that the result strongly depends on the constraints.
This property is reasonable because we will have 
an infinitely small duration if we do not impose any constraints.
Conversely, the result seems to be insensitive to the measure in Eq.~(\ref{T}).
It is reflected in the definition of $\delta_i$ in Eq.~(\ref{deltai}),  
but the explicit form is not important for the final BE (\ref{beq}).
These properties are the same as in the case of the quantum BE.

Generally, it is a difficult task to parametrize the final state. 
Here, we proceed as follows.
We set the parameters in $G_\perp$ to find an appropriate final state $p(T)$.
The transition-rate matrix $W$ is parametrized under the constraints (i)-(v), 
and the unknown parameters are obtained by solving the BE. 
This procedure corresponds to specifying the initial position and velocity, 
instead of fixing the initial and final positions 
in the variational method, as is the case for classical mechanics.

We also mention here the difference between the present BE and the quantum BE.
In the quantum case, we usually consider a time-dependent Hamiltonian, and 
the BE represents an equation of motion satisfied at all values of $t$.
Consequently, the optimized solution is characterized by 
a dynamical invariant~\cite{Takahashi,LR}.
This is not the case in the present system, and the corresponding invariant 
does not exist.
To clarify the meaning of the quantity $F$, 
we study the BE in Sec.~\ref{sec:sta} by allowing time-dependent fluctuations 
of the transition matrix $W$.

Below, we show how to solve the BE (\ref{beq}).
By using the eigenstates introduced in Sec.~\ref{sec:meq}, we write 
\be
 && |P(0)\rangle-|\pi\rangle = \sum_{i=1}^{N-1}p_i|R_i\rangle, \\
 && \langle\lambda_\perp|=\sum_{i=1}^{N-1}\lambda_i\langle L_i|.
\ee
The coefficients $p_i$ and $\lambda_i$ are determined from the initial condition.

Possible general forms of the operator $F$ are considered as follows.
As a constraint, we assume the form 
\be
 f_a = \Tr W X_a,
\ee
where $X_a$ represents a matrix.
For example, when we specify the $ji$ component of $W$, 
we use $(X)_{\mu\nu}=\delta_{\mu i}\delta_{\nu j}$.
Then $F$ is written as 
\be
 F = \sum_{a}\lambda^{({\rm v})}_aX_a.
\ee
By using the spectral decomposition, we write the BE as
\be
 \sum_a\lambda_a^{({\rm v})} \langle L_i|X_a|R_j\rangle
 = \frac{\e^{(\Lambda_i-\Lambda_j)T}-1}{\Lambda_i-\Lambda_j}
 p_i\lambda_j.
\ee
$W$ and $\lambda_a^{({\rm v})}$ are determined for given $p=(p_1,\dots,p_{N-1})$, 
$\lambda=(\lambda_1,\dots,\lambda_{N-1})$, and $T$.
In $N$-dimensional space, $W$ has $(N-1)^2$ degrees of freedom 
if we consider constraints (iii) and (iv). 
Constraint (v) does not change the number of undetermined parameters, 
because a multiplier constant is introduced for each constraint.

Some of their parameters are determined by considering the trace of the equation.
We have 
\be
 &&\sum_a\lambda_a^{({\rm v})} 
 \left(\Tr  X_a-\langle\pi|X_a|\pi\rangle
 \right) = T\sum_{i=1}^{N-1}p_i\lambda_i, \label{qbtr} \\
 &&\sum_a\lambda_a^{({\rm v})} 
 \Tr W^kX_a = T\sum_{i=1}^{N-1}\Lambda_i^kp_i\lambda_i,
 \label{qbk}
\ee
where $k$ is an integer. 
These relations determine $N-1$ parameters. 
For the remaining parameters, we cannot derive any general formula 
without explicitly calculating the eigenstates of $W$.
In the following, we study a simple system with $N=3$ 
to observe how the parameters are determined from the BE.

\section{Three-state system}
\label{sec:three}

\subsection{Parametrization}

In the case of $N=3$, $W$ is parametrized as
\begin{widetext}
\be
 W=\bmat{ccc}
 -(\pi_2 a+\pi_3 b) & \sqrt{\pi_1\pi_2}a-\sqrt{\pi_3}\delta & 
 \sqrt{\pi_1\pi_3}b+\sqrt{\pi_2}\delta \\
 \sqrt{\pi_2\pi_1}a+\sqrt{\pi_3}\delta & -(\pi_3 c+\pi_1 a) & 
 \sqrt{\pi_2\pi_3}c-\sqrt{\pi_1}\delta \\
 \sqrt{\pi_3\pi_1}b-\sqrt{\pi_2}\delta & 
 \sqrt{\pi_3\pi_2}c+\sqrt{\pi_1}\delta & -(\pi_1 b+\pi_2 c) \emat, 
\ee
\end{widetext}
where $a$, $b$, $c$, and $\delta$ are real parameters.
This parametrization is derived from constraints (iii) and (iv).
As the off-diagonal elements of $W$ represent probabilities, 
$a$, $b$, and $c$ must be non-negative.
Some of them are fixed by constraints, and the others are determined by the BE.
We note that nonzero values of $\delta$ 
break the transpose symmetry and signify violation of the DBC.
We also need the condition 
\be
 |\delta|\le {\rm Min}\left(
 \sqrt{\frac{\pi_1\pi_2}{\pi_3}}a, 
 \sqrt{\frac{\pi_3\pi_1}{\pi_2}}b, 
 \sqrt{\frac{\pi_2\pi_3}{\pi_1}}c\right), \label{nonnega}
\ee
which ensures non-negative values of the off-diagonal elements.

The initial distribution is parametrized as 
\be
 |P(0)\rangle=\bmat{c}
 p_1/\sqrt{\pi_1} \\
 p_2/\sqrt{\pi_2} \\
 p_3/\sqrt{\pi_3} 
 \emat,
\ee
where $p_{1,2,3}$ are non-negative numbers satisfying $p_1+p_2+p_3=1$.
In what follows, we consider the uniform distribution $p_1=p_2=p_3=1/3$.
It is difficult to parametrize the BE explicitly by using parameters 
in $|P(T)\rangle$, as mentioned in the previous section.
Therefore, we use a different parameter to characterize the BE.
It is naturally defined in the course of the following calculations.

The transition-rate matrix $W$ can be diagonalized as 
\be
 W = \Lambda_+|R_+\rangle\langle L_+|+\Lambda_-|R_-\rangle\langle L_-|,
\ee
where $|R_\pm\rangle$ and $\langle L_\pm|$ are the right and left eigenstates, 
respectively.
Their negative eigenvalues are given by 
\be
 && \Lambda_\pm = -\frac{A}{2}\pm\frac{\Delta}{2}.
\ee
We define 
\be
 && \Delta = \sqrt{A^2-4B}, \label{Delta}\\
 && A = (1-\pi_3)a+(1-\pi_2)b+(1-\pi_1)c, \\
 && B= \pi_1ab+\pi_3bc+\pi_2ca+\delta^2, \label{B}
\ee
where $\Delta$ represents the gap between two states $\pm$.
In the general formulation, $\Delta$ is not necessarily real, but 
the following analysis shows that it is real in the present three-state case. 
We also note that $A\ge 0$ and $B\ge 0$.
By using the eigenstates of $W$, 
we parametrize $|P(0)\rangle$ and $|\lambda_\perp\rangle$ as
\be
 && |P(0)\rangle-|\pi\rangle 
 = |R_+\rangle p_++|R_-\rangle p_-, \\
 && \langle\lambda_\perp| 
 = \lambda_+\langle L_+|
 +\lambda_-\langle L_-|.
\ee
The right-hand side of the BE is calculated as 
\be
 \int_0^{T} \diff t\,\e^{Wt}G_\perp \e^{-Wt} 
 = T\bmat{cc}
 p_+\lambda_+ & 
 p_+\lambda_-\langle\e\rangle \\
 p_-\lambda_+\langle\e^-\rangle & 
 p_-\lambda_- \emat, \no\\
\ee
where
\be
 && \langle\e^-\rangle 
 = \frac{1}{T}\int_0^{T} \diff t\,\e^{-\Delta t}
 = \frac{1-\e^{-\Delta T}}{\Delta T}, \\
 && \langle\e\rangle 
 = \frac{1}{T}\int_0^{T} \diff t\,\e^{\Delta t}
 = \frac{\e^{\Delta T}-1}{\Delta T}.
\ee

In the following analysis, we set $a=1$ as a constraint.
Then, the matrix $F$ is written as
\be
 F =\lambda^{({\rm v})} X
 = \lambda^{({\rm v})} \bmat{ccc} 0 & 1 & 0 \\ 1 & 0 & 0 \\ 0 & 0 & 0 \emat.
\ee
The BE reads
\be
 &&\lambda^{({\rm v})}
 \bmat{cc} \langle L_+|X|R_+\rangle & \langle L_+|X|R_-\rangle \\
 \langle L_-|X|R_+\rangle & \langle L_-|X|R_-\rangle \emat \no\\
 && =T\bmat{cc}
 p_+\lambda_+ & 
 p_+\lambda_-\langle\e\rangle \\
 p_-\lambda_+\langle\e^-\rangle & 
 p_-\lambda_- \emat. \label{2by2}
\ee

\subsection{Analysis of the brachistochrone equation}

To find the explicit form of Eq.~(\ref{2by2}), we first use Eq.~(\ref{qbtr}).
By taking the trace of Eq.~(\ref{2by2}), we obtain
\be 
 \frac{\lambda^{({\rm v})}}{T}
 =-\frac{p_+\lambda_++p_-\lambda_-}{2\sqrt{\pi_1\pi_2}}. \label{eq1}
\ee
This relation determines the multiplier $\lambda^{({\rm v})}$.
Next, we consider Eq.~(\ref{qbk}) with $k=1$, 
which gives 
\be
 A = 2a+\frac{1-z}{1+z}\Delta,
 \label{eq2}
\ee
where 
\be
 z = \frac{p_-\lambda_-}{p_+\lambda_+}. \label{z}
\ee
Equations (\ref{eq1}) and (\ref{eq2}) are obtained from 
the diagonal elements of the BE.

One of the remaining two equations is obtained by taking the trace of 
the square of Eq.~(\ref{2by2}).
After some calculations, we arrive at the expression
\be
 && \left(\frac{\sinh\frac{\Delta T}{2}}{\frac{\Delta T}{2}}\right)^2
 = 1+\frac{\pi_3}{\pi_1\pi_2}
 \frac{(1+z)^2}{4z}.
 \label{eq3}
\ee
The derived equation is parametrized by $z$, which is defined in Eq.~(\ref{z}).
We take this variable as a natural parameter characterizing 
the distribution at $t=0$.

To determine the last equation, we need to know the explicit form of 
the eigenfunctions $\langle L_\pm|$ and $|R_\pm\rangle$.
We consider the ratio of the off-diagonal parts of the BE, 
which is written as 
\be
 && \frac{\langle L_+|X|R_-\rangle}{\langle L_-|X|R_+\rangle} 
 = \frac{p_+\lambda_-}{p_-\lambda_+}\e^{\Delta T}.  \label{eq4-0}
\ee
The explicit form of this equation is discussed in what follows.

For given parameters $a$, $p_{1,2,3}$, $\pi_{1,2,3}$, and $z$, 
other parameters $b$, $c$, $\delta$, and $T$ are obtained 
from Eqs.~(\ref{eq2}), (\ref{eq3}), and (\ref{eq4-0}).
It is impossible to determine four parameters from three equations.
This uncertainty is resolved by demanding the minimum possible value of $T$.

\begin{center}
\begin{figure}[t]
\begin{center}
\includegraphics[width=0.9\columnwidth]{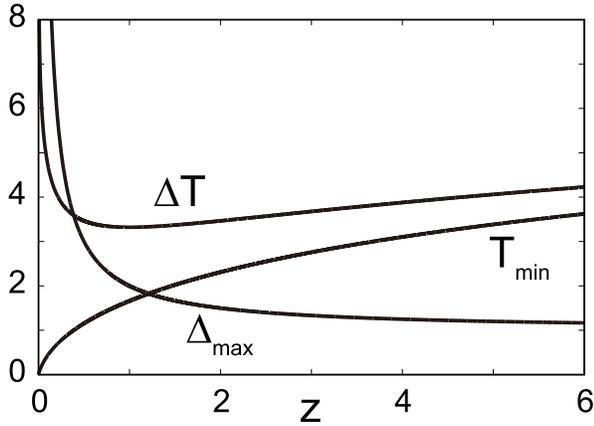}
\end{center}
\caption{Solution of $\Delta T$ in Eq.~(\ref{eq3}) with 
the maximum value of $\Delta$ and minimum of $T$ 
as a function of $z$.
We take $(\pi_1, \pi_2, \pi_3)=(0.5,0.3,0.2)$.
}
\label{fig-gap}
\end{figure}
\end{center}

The quantity $\Delta T$ is determined using Eq.~(\ref{eq3}).
As shown in Fig.~\ref{fig-gap}, we can determine a real solution for 
an arbitrary value of $z$ with $z\ge 0$.
To observe other possibilities, we consider the definition of $\Delta$ 
in Eq.~(\ref{Delta}).
We find that $\Delta$ can be purely imaginary when $A^2<4B$.
By inserting the expression $\Delta=i\tilde{\Delta}$ into Eq.~(\ref{eq3}), 
we see that the equation can be solved by using a negative $z$.
However, Eq.~(\ref{eq2}) shows that $(1-z)/(1+z)$ must be purely imaginary 
because $A$ is always real.
We put 
\be
 \frac{1-z}{1+z}=-i\alpha,
\ee
where $\alpha$ is real.
Equation (\ref{eq3}) is written as 
\be
 && \left(\frac{\sin\frac{\tilde{\Delta}T}{2}}{\frac{\tilde{\Delta}T}{2}}\right)^2
 = 1+\frac{\pi_3}{\pi_1\pi_2}\frac{1}{1+\alpha^2}.
\ee
This has no solution, and we conclude that 
the BE does not give complex eigenvalues in the present case.
This result is reasonable because the imaginary part of the eigenvalues of 
the transition-rate matrix does not accelerate relaxation.

The variable $A$ is written as Eq.~(\ref{eq2}) and $B=\frac{1}{4}(A^2-\Delta^2)$.
As these quantities must be non-negative, we have 
\be
 \Delta\le \frac{1+z}{z}a. \label{Dineq}
\ee
We seek the smallest $T$, 
which means that $\Delta$ takes the largest possible value.
When the equality holds in Eq.~(\ref{Dineq}), $B=0$, and 
we obtain a trivial solution $b=c=\delta=0$.
Therefore, to find the nontrivial result, we take a $\Delta$ 
smaller than the maximum possible value.
It is parametrized as 
\be
 \Delta = \frac{1+z}{z}ag, \label{g}
\ee
where $0\le g\le 1$.
Then  
\be
 A = a\left(2+\frac{1-z}{z}g\right). \label{Ag}
\ee
We also introduce $\tau$ with the condition $0\le \tau\le 1$ to write 
\be
 && b = \frac{A-(1-\pi_3)a}{1-\pi_2}\tau,  \label{bg}\\
 && c = \frac{A-(1-\pi_3)a}{1-\pi_1}(1-\tau). \label{cg}
\ee

Using these parametrizations,
we can rewrite Eq.~(\ref{eq4-0}) in the form 
\be 
 \delta = f(g,\tau,\delta^2,z), \label{eq4}
\ee
where $f$ is a function of the specified variables.
The detailed calculation and 
the explicit form of $f$ are presented in the Appendix.
Here, we simply mention that the parameters $p_\pm$ and $\lambda_\pm$ appear 
only through $z$.

For given $a$, $p_{1,2,3}$, and $\pi_{1,2,3}$, the solution 
is obtained as follows.
\begin{enumerate}
\item Fix the parameter $z$ with $z>0$.

\item Calculate $\Delta T$ from Eq.~(\ref{eq3}).

\item Take $\tau$ between 0 and 1.

\item Take $g$ between 0 and 1.

\item Calculate $A$, $\Delta$, $B$, $b$, $c$, and $\delta^2$ 
using Eqs.~(\ref{Delta}), (\ref{B}), (\ref{Ag}), (\ref{bg}), and (\ref{cg}). 

\item Check the positivity of the off-diagonal elements of $W$.
If not, go back to step 4 and 
repeat the calculation with a different $g$.

\item Calculate $\delta$ from Eq.~(\ref{eq4}).
Determine $g$ so that $\delta^2$ coincides with that calculated 
in Eq.~(\ref{B}).
Thus, $g$ is determined for a fixed $z$ and $\tau$.

\item Change $\tau$ and repeat the calculation.
Choose the solution with the largest $g$ ($\Delta$).
The parameters $g$ and $\tau$ are determined for a fixed $z$.

\item Obatin $T$ from $T=(\Delta T)/\Delta$.

\item Change $z$ and repeat the calculation. 

\end{enumerate}

\begin{center}
\begin{figure}[t]
\begin{center}
\includegraphics[width=0.9\columnwidth]{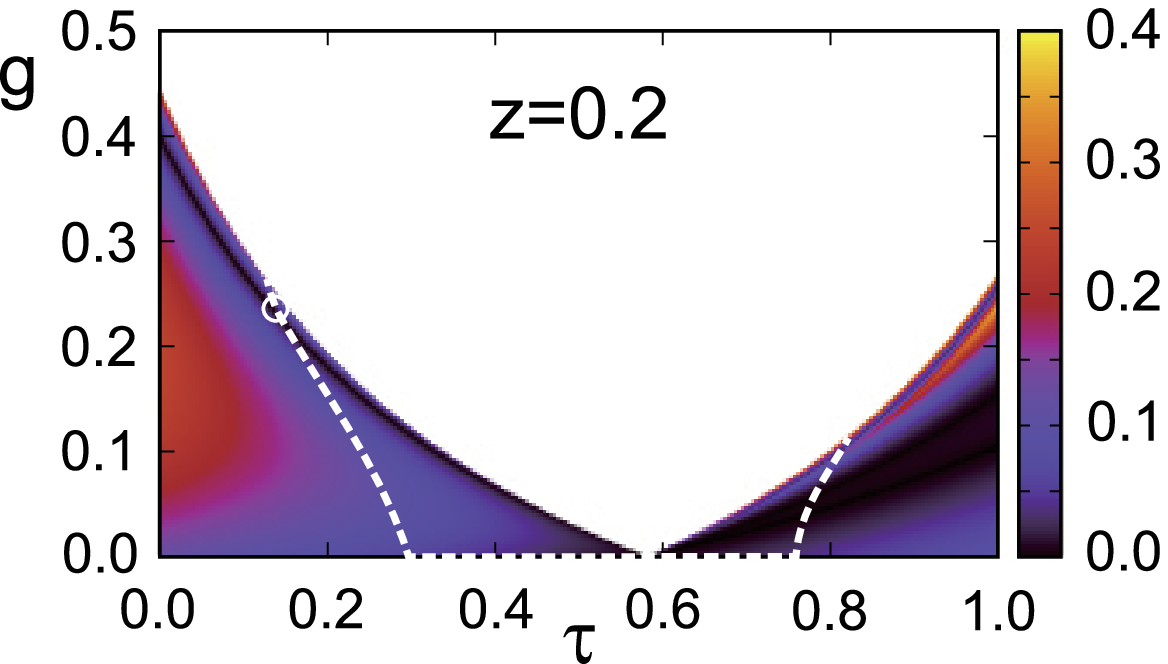}
\end{center}
\vspace{2mm}
\begin{center}
\includegraphics[width=0.9\columnwidth]{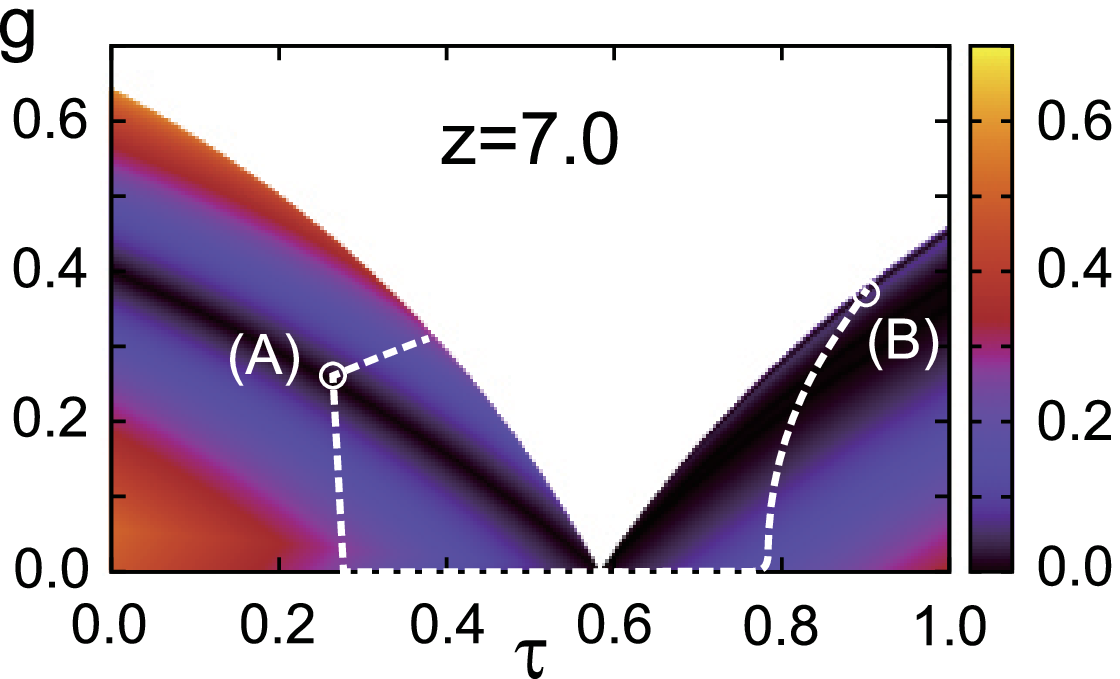}
\end{center}
\caption{$|\delta-f(g,\tau,\delta^2,z)|$ (Eq.~(\ref{eq4})) 
as a function of $g$ and $\tau$.
Top: The result with $z=0.2$. Bottom: The result with $z=7.0$.
Points satisfying condition (\ref{nonnega}) are specified 
by the region surrounded by the dashed curves.
For $z=0.2$, the optimal solution is denoted by the circle.
For $z=7.0$, the solution is denoted by point (B). 
Point (A) is used as a suboptimal solution in the following analysis.}
\label{deltadiff}
\end{figure}
\end{center}

\begin{center}
\begin{figure}[t]
\begin{center}
\includegraphics[width=0.9\columnwidth]{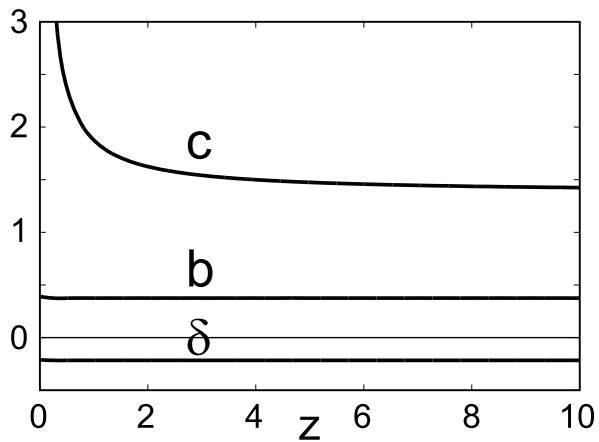}
\end{center}
\caption{Parameters $b$, $c$, and $\delta$ 
in the transition-rate matrix $W$. 
We take the solution with $b<c$.}
\label{bcd}
\end{figure}
\end{center}
\begin{center}
\begin{figure}[t]
\begin{center}
\includegraphics[width=0.9\columnwidth]{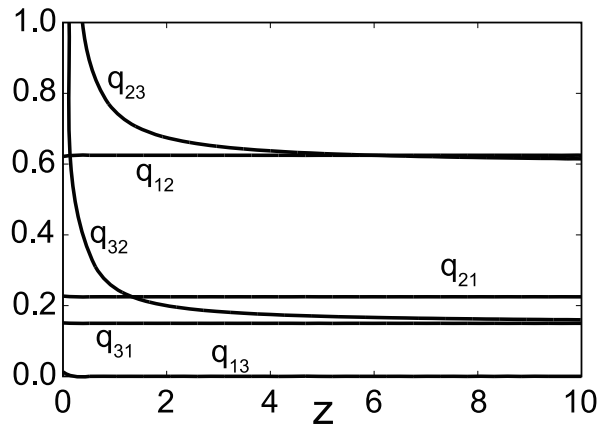}
\end{center}
\caption{Off-diagonal elements of $q$ representing transition rates.}
\label{q}
\end{figure}
\end{center}
\begin{center}
\begin{figure}[t]
\begin{center}
\includegraphics[width=0.9\columnwidth]{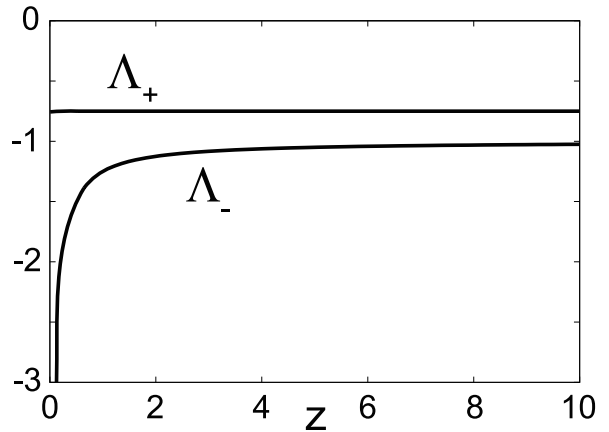}
\end{center}
\caption{Eigenvalues of matrix $W$.}
\label{ev}
\end{figure}
\end{center}
\begin{center}
\begin{figure}[t]
\begin{center}
\includegraphics[width=0.9\columnwidth]{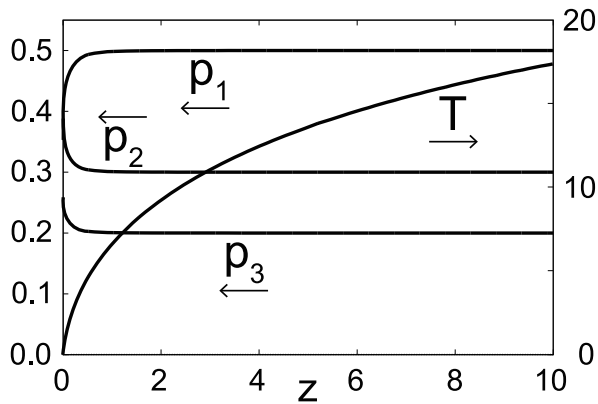}
\end{center}
\caption{Duration $T$ (rightmost grid line) and 
probability distribution function $p=(p_1, p_2, p_3)$ 
at $t=T$ (leftward grid lines).}
\label{ptf-tf}
\end{figure}
\end{center}

\begin{center}
\begin{figure}[t]
\begin{center}
\includegraphics[width=0.8\columnwidth]{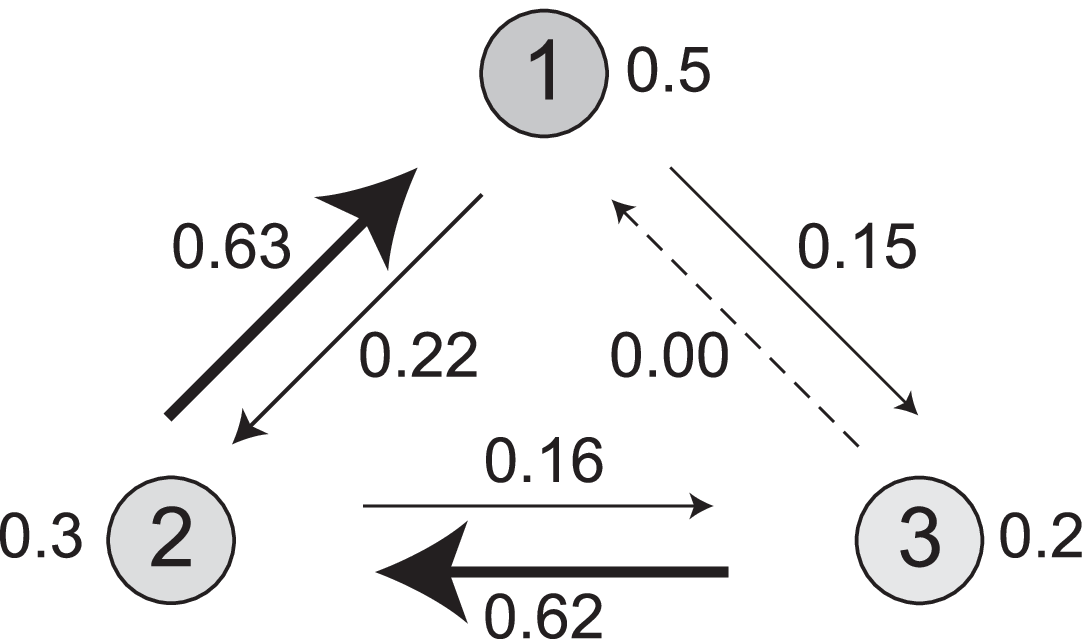}
\end{center}
\vspace{2mm}
\begin{center}
\includegraphics[width=0.8\columnwidth]{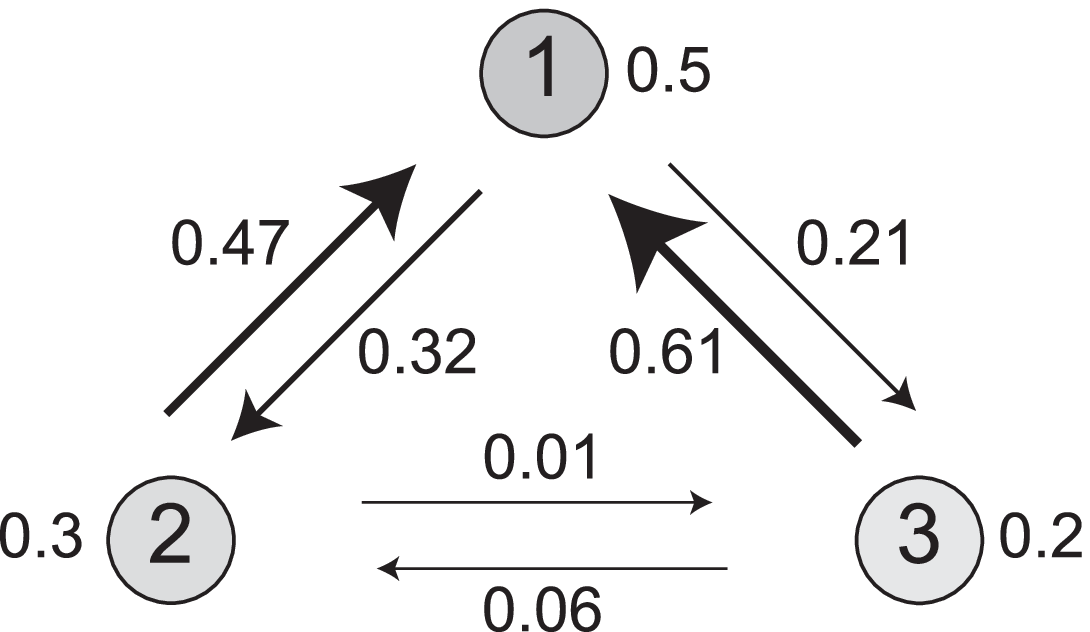}
\end{center}
\caption{Flow diagrams for $z=7.0$.
Top: Solution (A) in Fig.~\ref{deltadiff}.
Bottom: Solution (B) in Fig.~\ref{deltadiff}.
The width of the arrow from state $j$ to $i$ 
indicates the transition rate $q_{ij}$, which is also specified 
by the corresponding number.}
\label{q-z}
\end{figure}
\end{center}

We numerically solve the BE according to this algorithm.
In what follows, we set parameters $a=1$ and $(\pi_1,\pi_2,\pi_3)=(0.5,0.3,0.2)$.

To find the solution of Eq.~(\ref{eq4}), we plot $|\delta-f(g,\tau,\delta^2,z)|$ 
as a function of $g$ and $\tau$ for a fixed $z$ in Fig.~\ref{deltadiff}.
The result is plotted only when the $\delta^2$ obtained from Eq.~(\ref{B}) is positive.
We also need to impose the condition of non-negativity of the off-diagonal elements
of $q$ in Eq.~(\ref{nonnega}). 
Equation (\ref{eq4}) has many solutions, and we select, in principle, 
the one with the largest value of $g$.
We see from Fig.~\ref{deltadiff} that the solutions are classified into 
left and right branches.
The solutions in the left branch are aligned along a single curve, and 
those in the right branch appear to be more complicated.
We consider the left branch in the following calculations, and 
the meaning of the branches is discussed afterwards.

By taking the optimum $g$ and $\tau$, 
we can obtain $\delta$ and calculate other parameters.
In Fig.~\ref{bcd}, we plot the result of $b$, $c$, and $\delta$ for a given $z$.
Nonzero values of $\delta$ show violation of the DBC.
We expect from the previous analysis that a larger $\delta$ accelerates 
relaxation~\cite{IO1}.
However, each matrix element of $q$ must be non-negative.
As we see in Fig.~\ref{q}, where all matrix elements are plotted, 
one of the elements $q_{13}$ is almost equal to zero.
This means that $|\delta|$ takes the maximum possible value, and 
the violation of DBC is maximal.

We also show the nonzero eigenvalues of $W$ in Fig.~\ref{ev}, and 
the duration $T$ and probability distribution at $t=T$ 
in Fig.~\ref{ptf-tf}.
If the value of $z$ is not so small, 
the probability distribution $p(T)$ is very close to 
the stationary distribution $\pi$ at $t\to\infty$.
This is a practically useful property because we usually do not set 
the distribution at a finite duration.

As mentioned above, the solutions of the BE are classified into two parts.
To determine the meaning of the solutions, we take the solutions 
in both parts for $z=7.0$, which are shown in Fig.~\ref{q-z}.
Solutions (A) and (B) represent the points in Fig.~\ref{deltadiff}, respectively.
We see that solution (A) in the left branch gives a flow $3\to 2\to 1$, and 
the direct transition between 1 and 3 is suppressed.
This is a reasonable result because the stationary distribution 
in the present calculation satisfies $\pi_3>\pi_2>\pi_1$.
On the other hand, the right branch satisfies $b>c$, and 
the transition between 2 and 3 is suppressed.
In other words, the flow is not in a single direction and becomes complicated.
This is considered to be the reason why the solutions of the right branch 
in Fig.~\ref{deltadiff} are more complicated than those of the left branch.

\subsection{Optimality of the solution}

\begin{center}
\begin{figure}[t]
\begin{center}
\includegraphics[width=0.9\columnwidth]{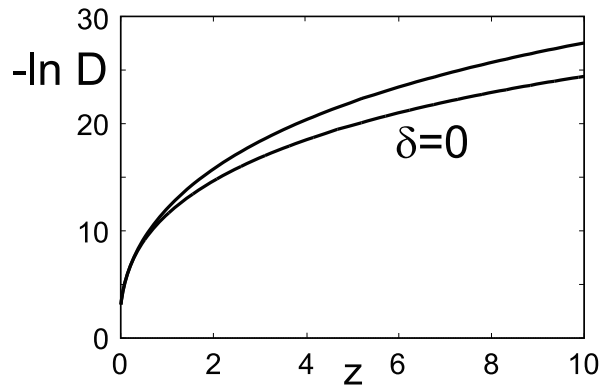}
\end{center}
\caption{Kullback--Leibler divergence $-\ln D(p(T;\delta)|\pi)$ 
and $-\ln D(p(T;\delta=0)|\pi)$.}
\label{kl-kl0}
\end{figure}
\end{center}

To see the optimality of the solution, we calculate the Kullback--Leibler divergence 
$D(p(T)|\pi)$ between state $p(T)$ and $\pi$.
This value is compared with the same quantity with $\delta=0$ and 
the other parameters unchanged.
The result is plotted in Fig.~\ref{kl-kl0} and 
shows that $D(p(T;\delta)|\pi)<D(p(T;0)|\pi)$.
The inclusion of the violation parameter indeed accelerates 
relaxation, as discussed in Ref.~\cite{IO1}.

In principle, our optimization is performed to minimize the duration between 
$p(0)$ and $p(T)$.
Although $p(T)$ is very close to the stationary distribution $\pi$, 
as we show in Fig.~\ref{ptf-tf}, the optimality of the approach to $\pi$ 
is not guaranteed. 
Therefore, the use of $D(p(T)|\pi)$ is not justified to prove 
the optimality of the solution.

\begin{center}
\begin{figure}[t]
\begin{center}
\includegraphics[width=0.9\columnwidth]{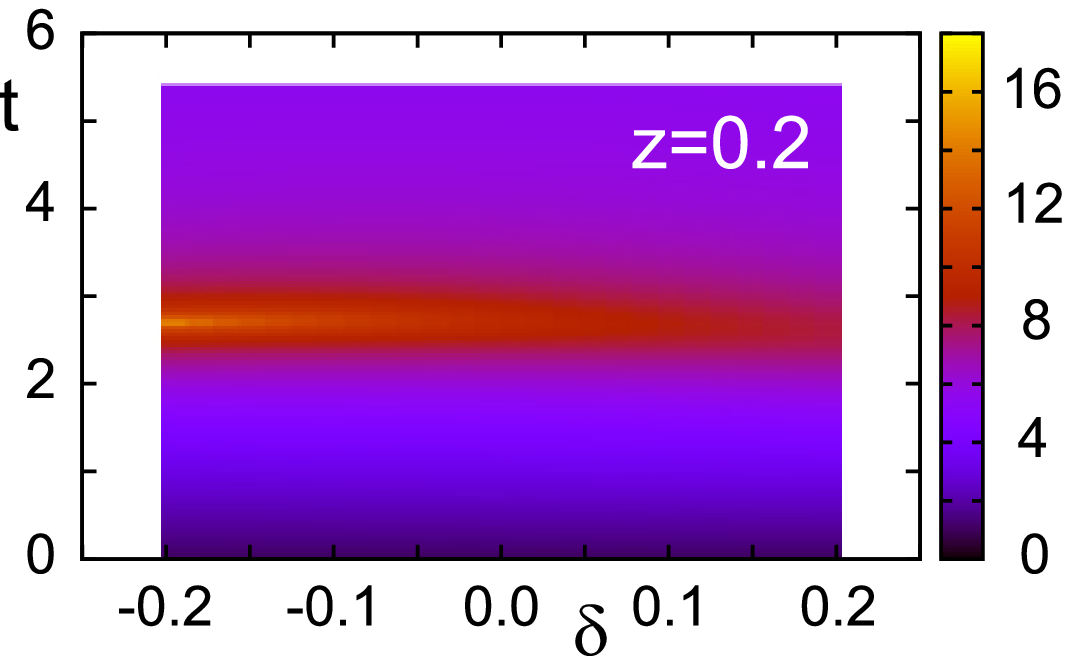}
\end{center}
\vspace{2mm}
\begin{center}
\includegraphics[width=0.9\columnwidth]{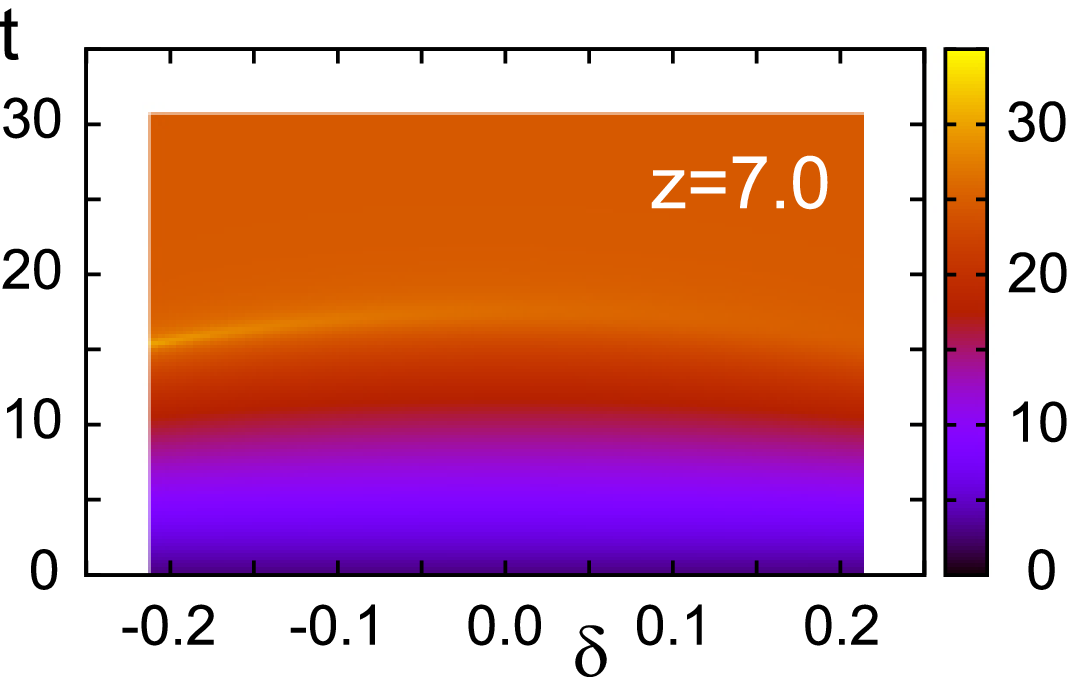}
\end{center}
\caption{$-\ln D(p(t,\delta)|p_{\rm opt})$ as a function of $t$ and $\delta$.
Top: The result with $z=0.2$. 
Bottom: The result with $z=7.0$.}
\label{kl-z}
\end{figure}
\end{center}

We consider the distance between the final distribution $p_{\rm opt}=p(T)$ and 
the distribution $p(t,\delta)$ as a function of $t$ and $\delta$.
The other parameters $a$, $b$, and $c$ are the same as those in $p_{\rm opt}$.
We change $\delta$ within the region where $\Delta$ becomes real.
That is, we have
\be
 && \delta^2\le \frac{1}{4}(A^2-4B_0), \\
 && B_0=\pi_1ab+\pi_2ca+\pi_3bc.
\ee
We also require that $W_{ij}\ge 0$ ($i\ne j$).
The result is plotted in Fig.~\ref{kl-z}.
The color map represents $-\ln D(p(t,\delta)|p_{\rm opt})$ 
as a function of $t$ and $\delta$.
We see that the solution of the BE takes the minimum value 
of $D(p(t,\delta)|p_{\rm opt})$, which indeed represents the optimality 
of the solution.

\begin{center}
\begin{figure}[t]
\begin{center}
\includegraphics[width=0.85\columnwidth]{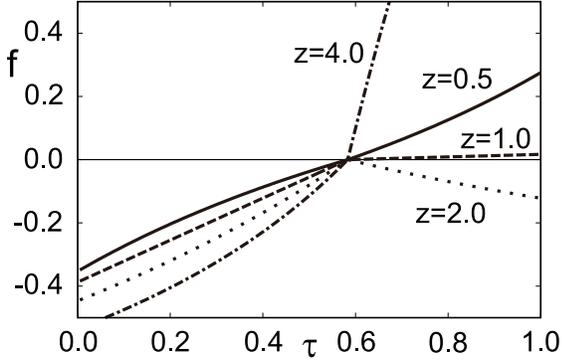}
\end{center}
\caption{$f(g(\tau),\tau,0,z)$ in Eq.~(\ref{eq4}) with $\delta=0$.}
\label{delta0}
\end{figure}
\end{center}

Further confirmation of the optimization is obtained by solving the BE within the DBC.
In this case, we set $\delta=0$, and $g$ and 
$\tau$ are related to each other by Eq.~(\ref{B}), 
which is a quadratic equation that is easily solved to obtain $g=g(\tau)$.
In Fig.~\ref{delta0}, we plot the function $f(g(\tau),\tau,\delta,z)$ on 
the right-hand side of Eq.~(\ref{eq4}).
The solution is given by the value of $\tau$ satisfying $f=0$.
We see that it is independent of $z$. 
By using the relation between $g$ and $\tau$, we obtain $g=0$.
This solution represents the boundary point 
between the left and right branches in Fig.~\ref{deltadiff}.
We have that $\Delta=0$ and $T$ tends to $\infty$.
Thus, the BE with the DBC does not give any solution.
This means that we cannot reach the final state determined by $z$ 
with a finite duration in the algorithm with DBC.

\subsection{Optimization and detailed balance condition}

In the above example, we have studied the case $a=1$.
We find that one of the elements of the transition-rate matrix 
tends to 0, which means that the violation of the DBC is maximal.
We further investigate whether it is a general property of the solution.
To answer this question, we take the simple case $a=b=c$.
Then, the constraint matrix is given by
\be
 F = \sum_{a=1}^3 \lambda_a^{({\rm v})}X_a,
\ee
where $\sum_{a=1}^3 \lambda_a^{({\rm v})}=0$, 
$(X_1)_{ij}=\delta_{i1}\delta_{j2}+\delta_{i2}\delta_{j1}$, 
$(X_2)_{ij}=\delta_{i1}\delta_{j3}+\delta_{i3}\delta_{j1}$, and 
$(X_3)_{ij}=\delta_{i2}\delta_{j3}+\delta_{i3}\delta_{j2}$.
The eigenvalues of $W$ are given by 
\be
 \Lambda_\pm = -a\pm\sqrt{-\delta^2}.
\ee
Then, Eq.~(\ref{eq1}) is replaced by
\be
 && -2\left(\sqrt{\pi_1\pi_2}\lambda_1^{({\rm v})}
 +\sqrt{\pi_1\pi_3}\lambda_2^{({\rm v})}
 +\sqrt{\pi_2\pi_3}\lambda_3^{({\rm v})}\right) \no\\
 && =T(p_+\lambda_++p_-\lambda_-),
\ee
and Eq.~(\ref{eq2}) by  
\be
 \frac{1-z}{1+z}\Delta=\frac{1-z}{1+z}\sqrt{-4\delta^2}=0.
\ee
The latter equation implies $\delta=0$.
This means that the optimized solution in the case with the constraint $a=b=c$
does not violate the DBC.

This example may be too simple, but it is instructive to find 
a solution respecting the DBC.
For example, if we specify $a$, $b$, and $c$ with $a=b>c$, 
the result becomes different and we can have nonzero $\delta$.
Thus, the result is highly sensitive to the constraints, as mentioned above, 
and the optimized solution does not necessarily violate the DBC.
This is not a surprising result, because the previous study of 
the optimization to reduce the rejection rate 
in the Markov-chain Monte Carlo method yielded 
a trivial solution satisfying the DBC in a simple case~\cite{ST}.

\section{Shortcuts to adiabaticity}
\label{sec:sta}

In our BE, the matrix $F$ constructed from constraint functions 
as Eq.~(\ref{F}) plays an important role to describe the optimized solution.
However, the physical meaning of $F$ was unclear in the above analysis.
This is considered to be due to the strong restriction of 
the transition-rate matrix $W$.
If we allow time-dependent fluctuations of $W$, 
we expect from the analysis of the quantum BE that 
$F$ represents the dynamical invariant~\cite{Takahashi}.
We can, in principle, formulate the BE for nonstatic systems.
By examining the problem from a more general perspective, 
we can learn the general properties of the solution.
We can also expect that the present analysis is applied to 
simulated annealing~\cite{KGV}.

When $W$ has time dependence, all the constraints are locally imposed in time.
We modify the action in a local form by using the Lagrangian $L$ 
as $S=\int \diff t\, L$.
The constraint part of the Lagrangian is given by 
\be
 L_{\rm c} &=& \sum_{i=1}^N\lambda_i^{({\rm i})}(t)\Biggl(
 \frac{\diff}{\diff t}P_i(t)-\sum_{j=1}^NW_{ij}(t)P_j(t)\Biggr) \no\\
 && +\lambda^{({\rm ii})}(t)\left(1-\sum_{i=1}^N
 \sqrt{\pi_i}P_{i}(t)\right) \no\\
 && +\sum_{i,j=1}^N\left(\sqrt{\pi_i}W_{ij}(t)\lambda_j^{({\rm iii})}(t)
 +\lambda_i^{({\rm iv})}(t)W_{ij}(t)\sqrt{\pi_j}\right) \no\\
 && +\sum_a\lambda_a^{({\rm v})}(t) f_a(W(t)). 
\ee
Each constraint is the same as in the time-independent case.
We note that constraint (iv) is represented as
\be
 \sum_{j=1}^N W_{ij}(t)\sqrt{\pi}_j = 0. \label{dbct}
\ee 
In principle, the balance condition is expressed as 
$\sum_j W_{ij}(\infty)\sqrt{\pi}_j = 0$, and it is not necessary 
to impose the balance condition at each $t$.
However, we find that it is difficult to construct a general solution 
if we do not impose condition (\ref{dbct}).
In what follows, we consider this special case with Eq.~(\ref{dbct}).

The variational procedure is the same as in the time independent case.
We finally obtain the form 
\be
 F_\perp(t)=U(t)G_\perp U^{-1}(t), \label{fd}
\ee
where $U$ is the time-evolution operator: 
\be
 U(t)={\rm T}\exp\left(\int^t_0\diff t'\,W(t')\right).
\ee
The symbol T denotes the time-ordered product.
The definitions of $G_\perp$ and $F_\perp$ are the same as before.
We see that this result (\ref{fd}) is a natural extension of Eq.~(\ref{beq}).
The difference from the previous case is that the constraint matrix $F$ 
is time-dependent and we can consider the time derivative of $F$,
which satisfies the equation of motion 
\be
 \frac{\diff F_\perp(t)}{\diff t} = [W(t), F_\perp(t)]. \label{eqm}
\ee
In this case, $F$ has a clear meaning: it is a dynamical invariant~\cite{LR}.
We consider the symmetric case $F=F^{\rm T}$.
This means that the constraints are imposed only on the part with DBC.
In this case, the matrix is diagonalized as 
\be
 F_\perp(t)=\sum_{n}f_n |n(t)\rangle\langle n(t)|.
\ee
We can easily show from Eq.~(\ref{eqm}) that 
the eigenvalues $f_n$ are time independent.
By using the eigenstates, we can also show that 
\be
 \langle m(t)|W(t)|n(t)\rangle = \langle m(t)|\dot{n}(t)\rangle,
\ee
where $m\ne n$.
This relation means that $W$ is written as 
\be
 W(t) &=& \sum_n w_n(t)|n(t)\rangle\langle n(t)| \no\\
 && +\sum_{m\ne n}|m(t)\rangle\langle m(t)|\dot{n}(t)\rangle\langle n(t)| \no\\
 &=& \sum_n \left(w_n(t)-\langle n(t)|\dot{n}(t)\rangle\right)
 |n(t)\rangle\langle n(t)| \no\\
 && +\sum_n|\dot{n}(t)\rangle\langle n(t)|, \label{wsta}
\ee
where $w_n(t)$ is a function of $t$ and cannot be determined 
from Eq.~(\ref{eqm}).
We note that the second term in Eq.~(\ref{wsta}) is asymmetric:
\be
 \left(\sum_n|\dot{n}(t)\rangle\langle n(t)|\right)^{\rm T}
 = -\sum_n|\dot{n}(t)\rangle\langle n(t)|.
\ee
This property shows that the optimal transition-rate matrix violates the DBC.

The present result is interpreted as the counterdiabatic driving.
The probability distribution is given by 
\be
 |P(t)\rangle = \sum_n |n(t)\rangle
 \exp\left(\int_0^t \diff t'\,w_n(t')\right)+|\pi\rangle.
\ee
This means that the probability distribution 
is given by the instantaneous eigenstate of the first term in Eq.~(\ref{wsta}).
The second term represents the counterdiabatic term
and prevents nonadiabatic transitions to different states.
The counterdiabatic term explicitly violates the DBC.
This is the main result of the present section.
In quantum systems, the counterdiabatic term usually takes 
an operator that is absent in the original Hamiltonian.
The same is true for the Markovian dynamics.
We need to break the DBC to prevent non-adiabatic transitions.

\section{Discussion and conclusions}
\label{sec:sum}

We performed optimization for Markovian dynamics 
through the brachistochrone method.
The master equation, which represents the classical stochastic dynamics, 
has a formal similarity to the Schr\"odinger equation in quantum systems.
Therefore, we applied the formalism of the quantum brachistochrone method 
to our case by replacing the measure with a proper one.
We derived the BE and showed that the optimized solution can violate the DBC.
We took a simple case with three states to demonstrate our method and 
various properties of the optimal solution in detail.
We also found that the optimized solution was interpreted as 
the counterdiabatic driving by considering the time-dependent transition-rate matrix.
In this case, the solution is characterized by the dynamical invariant.
The probability distribution follows an adiabatic passage, and 
the counterdiabatic term in the transition-rate matrix violates the DBC.

Our result shows that the optimized solution significantly depends on 
the constraints to be used.
A part of the optimized solution satisfies the DBC, but in general, the optimized solution violates the DBC.
This means that the violation of the DBC does not necessarily optimize the time evolution.
The same property was obtained by Suwa and Todo in Ref.~\cite{ST},
where they optimized the stochastic dynamics with respect to the rejection rate
and found that the DBC is not violated in some simple cases. 
On the other hand, if we allow time dependence in the transition-rate matrix, 
the general solution violates the DBC.
The violation originates from the counterdiabatic term to keep the system 
in the instantaneous stationary state.
In other words, violation of the DBC immediately relaxes the system 
to the instantaneous stationary state in the time-dependent case.

The starting point of our analysis is to use the Kullback--Leibler divergence 
as the measure of optimization, and we pursue the ``fastest" solution.
However, by using the defined action, we found that the general form of the BE 
is not sensitive to the form of the measure, as in the case of quantum systems.
This property is very convenient to us because we can avoid the uncertainty caused by the fact that 
the meaning of ``fastest'' depends on the choice of measure. 
There exists a geometrical interpretation 
in the quantum brachistochrone method~\cite{NDGD}.
We expect that a similar interpretation holds in the present case of 
classical stochastic dynamics.

Originally, we aimed at the optimal solution leading to 
the fastest convergence to the stationary distribution.
Although the stationary state can be achieved after infinite-time relaxation, 
in our formulation, the optimization is performed between two states 
at a finite duration.
In this sense, we do not directly attain the optimal pathway 
between the initial and stationary states.
This is due to the limitation of the original quantum brachistochrone method.
Thus, we focused not on the relaxation time but on the finite duration.
Nevertheless, in this formulation, 
we found that the optimal solution violates the DBC 
to reach the specified final distribution.
Furthermore, in three-state systems, the distribution $p(T)$ at the final time 
is very close to the stationary distribution $\pi$, 
which means that our optimization practically makes sense 
in finding the optimal solution to the stationary state.
We hope that future studies fulfill the gap in the formulation.

Considering the above properties of our analysis, 
we offer possible applications of our method.
Certainly, it will be difficult to optimize the dynamics fully in large systems.
However, it is possible to optimize a portion of the elements 
in the transition-rate matrix by imposing several constraints. 
In principle, we can implement any constraint in our method.
This flexibility is one of the advantages of our approach and 
will be useful for practical applications.

Before closing the last section, we list below possible directions for 
future research in the subject of the present study.
The adiabatic passage is translated into a quasistatic process 
to maintain system equilibrium in the context of the classical counterpart.
The time-dependent driving appears in the cases of simulated annealing 
to find an optimal solution in the rugged landscape of the energy~\cite{KGV} 
and in the case of machine learning to estimate optimal parameters characterizing 
the stationary distribution by driving the stochastic dynamics 
many times~\cite{WH,SBD,Ohzeki}.
We can develop a more efficient algorithm for these applications 
by utilizing the brachistochrone method for the classical stochastic dynamics, 
as in our study.
Actually, the dynamics modified by violating the DBC leads to 
an efficient algorithm in learning~\cite{Ohzeki}.

As shown in our study, use of the BE can be a useful method 
to design the transition-rate matrix.
It is straightforward to generalize our scheme to the case of 
the Fokker--Planck equation, which is also interpreted 
as the imaginary-time Schr\"odinger equation.
We naturally expect that the BE leads to the special force 
to accelerate convergence to the stationary state, 
as proposed in Refs.~\cite{OI,OI2}, and beyond.
Finally, future studies may expand the present analysis 
to the case of discrete-time evolution by changing 
the formulation of the quantum brachistochrone because 
the implementation in numerical computations can be realized in discrete time.
In this sense, the present study is merely a starting point 
to construct the best algorithm simulating the optimal stochastic dynamics.
After various studies on the subject of the present study, 
we hope that a type of constructive concept to generate 
an algorithm more efficient than DBC emerges in the future.

\section*{Acknowledgments}

The authors are grateful to A. Ichiki for useful comments.
KT was supported by JSPS KAKENHI Grant Number 26400385.
MO was supported by JSPS KAKENHI Grant Number 15H03699 
and the Kayamori Foundation of Informational Science Advancement.
The authors are grateful for financial support from the JSPS Core-to-Core program, 
Non-equilibrium Dynamics of Soft Matter and Information.

\appendix
\section{Derivation of the brachistochrone equation}

The purpose of this Appendix is 
to present the explicit form of Eq.~(\ref{eq4-0}).
For this purpose, we need eigenstates of $W$ with nonzero eigenvalues: 
\be
 &&\langle L_\pm| = \bmat{ccc} \alpha'_\pm & \beta'_\pm & \gamma'_\pm \emat, \\
 &&|R_\pm\rangle = \bmat{c} \alpha_\pm \\ \beta_\pm \\ \gamma_\pm \emat.
\ee
Using $\langle\pi|L_\pm\rangle=0$ and 
$\langle R_\pm|\pi\rangle=0$, we have 
\be
 && \gamma = 
 -\sqrt{\frac{\pi_1}{\pi_3}}\alpha
 -\sqrt{\frac{\pi_2}{\pi_3}}\beta,\\
 && \gamma' = 
 -\sqrt{\frac{\pi_1}{\pi_3}}\alpha'
 -\sqrt{\frac{\pi_2}{\pi_3}}\beta'.
\ee
The eigenvalue equations
$W|R\rangle=\Lambda |R\rangle$
and $\langle L|W= \langle L|\Lambda$ give 
\be
 && \beta = -C\alpha=-\left(
 \frac{W_{11}-\sqrt{\frac{\pi_1}{\pi_3}}W_{13}-\Lambda}
 {W_{12}-\sqrt{\frac{\pi_2}{\pi_3}}W_{13}}\right)\alpha,
 \\
 && \beta' =-C'\alpha'= -\left(
 \frac{W_{11}-\sqrt{\frac{\pi_1}{\pi_3}}W_{13}-\Lambda}
 {W_{21}-\sqrt{\frac{\pi_2}{\pi_3}}W_{31}}\right)\alpha'.
\ee
These conditions determine the relations between components of the eigenstate vectors.
The left and right eigenstates are related by 
the normalization condition $\langle L_i|R_i\rangle=1$.
We obtain 
\be
 \frac{1}{\pi_3}\Bigl[1-\pi_2
 -\sqrt{\pi_1\pi_2}\left(C+C'\right)+(1-\pi_1)CC'
 \Bigr]\alpha\alpha'=1.  \no\\
\ee
We have omitted the subscript $\pm$ 
in the above equations.
The normalization condition does not fully determine 
$\alpha$ and $\alpha'$.
The substitutions $\alpha\to k\alpha$ and $\alpha'\to \alpha'/k$
do not change the condition.

Now, we return to Eq.~(\ref{eq4-0}).
Using the obtained relations of the eigenstate vectors, we can write 
\be
 && \langle L_+|X|R_-\rangle
 = -\left(C_-+C'_+\right)\alpha'_+\alpha_-, \\
 && \langle L_-|X|R_+\rangle
 = -\left(C_++C'_-\right)\alpha'_-\alpha_+.
\ee
Then, the ratio is written as 
\begin{widetext}
\be
 && \frac{C_-+C'_+}{C_++C'_-}
 \frac{1-\pi_2
 -\sqrt{\pi_1\pi_2}\left(C_++C'_+\right)+(1-\pi_1)C_+C'_+}{1-\pi_2
 -\sqrt{\pi_1\pi_2}\left(C_-+C'_-\right)+(1-\pi_1)C_-C'_-}
 \left(\frac{\alpha'_+}{\alpha'_-}\right)^2
 = \frac{p_+\phi_-}{p_-\phi_+}\e^{\Delta T}. \label{a2}
\ee

Next, we consider 
the relations between $p_\pm$ and $\alpha'_\pm$.
They are determined by $p_\pm =\langle L_\pm|P(0)\rangle$.
We obtain
\be
 p_\pm=\frac{1}{\pi_3}\left[
 \frac{1}{\sqrt{\pi_1}}(p_1\pi_3-p_3\pi_1)
 -\frac{1}{\sqrt{\pi_2}}(p_2\pi_3-p_3\pi_2)C'_\pm
 \right]\alpha'_\pm. 
\ee
These are inserted to Eq.~(\ref{a2}) to obtain
\be
 && \frac{C_-+C'_+}{C_++C'_-}
 \left[\frac{1-\pi_2
 -\sqrt{\pi_1\pi_2}\left(C_++C'_+\right)+(1-\pi_1)C_+C'_+}{1-\pi_2
 -\sqrt{\pi_1\pi_2}\left(C_-+C'_-\right)+(1-\pi_1)C_-C'_-}\right]
 \left[\frac{\frac{1}{\sqrt{\pi_1}}(p_1\pi_3-p_3\pi_1)
 -\frac{1}{\sqrt{\pi_2}}(p_2\pi_3-p_3\pi_2)C'_-}
 {\frac{1}{\sqrt{\pi_1}}(p_1\pi_3-p_3\pi_1)
 -\frac{1}{\sqrt{\pi_2}}(p_2\pi_3-p_3\pi_2)C'_+}\right]^2
 = z\e^{\Delta T}. \no\\
\ee
We see that the arbitrariness of the choice of 
eigenfunctions does not appear in this expression, as expected.
The last equation is deformed as 
\be
 && \delta = -\frac{D_+z\e^{\Delta T}\left(D\frac{D_+^2+\delta^2}{2D_+}
 -\delta^2\right)-D_-D_C\left(D\frac{D_-^2+\delta^2}{2D_-}+\delta^2\right)}
 {D_+z\e^{\Delta T}\left(D-\frac{D_+^2+\delta^2}{2D_+}\right)
 -D_-D_C\left(D+\frac{D_-^2+\delta^2}{2D_-}\right)}, \label{eq4-e}
\ee
where
\be
 && D=\frac{2}{\Delta}\sqrt{\frac{\pi_1\pi_2}{\pi_3}}
 \left[\frac{\pi_3(a-b)}{2(1-\pi_1)}
 \Bigl(2(1-\pi_2)(a-b)+A-2a\Bigr)
 -\delta^2\right], \\
 && D_\pm=\frac{\frac{1}{\sqrt{\pi_1}}(p_1\pi_3-p_3\pi_1)\sqrt{\pi_1\pi_2}(a-b)
 -\frac{1}{\sqrt{\pi_2}}(p_2\pi_3-p_3\pi_2)
 \Bigl((1-\pi_2)(a-b)-(\Lambda_\pm+a)\Bigr)
 }
 {\frac{1}{\sqrt{\pi_1}}(p_1\pi_3-p_3\pi_1)\frac{1-\pi_1}{\sqrt{\pi_3}}
 -\frac{1}{\sqrt{\pi_2}}(p_2\pi_3-p_3\pi_2)\sqrt{\frac{\pi_1\pi_2}{\pi_3}}}, \\
 && D_C=
 \frac{1-\pi_2
 -\sqrt{\pi_1\pi_2}\left(C_++C'_+\right)+(1-\pi_1)C_+C'_+}{1-\pi_2
 -\sqrt{\pi_1\pi_2}\left(C_-+C'_-\right)+(1-\pi_1)C_-C'_-}.
\ee
\end{widetext}
We conclude that the explicit form of Eq.~(\ref{eq4})
is given by Eq.~(\ref{eq4-e}).
We note that the right-hand side of (\ref{eq4-e}) 
depends on $\delta^2$, rather than on $\delta$.
This fact is shown by using the property $D_C(-\delta)=D_C(\delta)$.



\section*{References}
\begin{thebibliography}{99}

\bibitem{MRRTT}
N. Metropolis, A. W. Rosenbluth, M. N. Rosenbluth, A. H. Teller, 
and E. Teller, J. Chem. Phys. {\bf 21}, 1087 (1953).

\bibitem{Hastings}
W. K. Hastings, Biometrika {\bf 57}, 97 (1970). 

\bibitem{ST}
H. Suwa and S. Todo, Phys. Rev. Lett. {\bf 105}, 120603 (2010).

\bibitem{TCV}
K. S. Turitsyn, M. Chertkov, and M. Vucelja, Physica D {\bf 240}, 410 (2011).

\bibitem{FW}
H. C. M. Fernandes and M. Weigel, Comput. Phys. {\bf 182}, 1856 (2011).

\bibitem{SH}
Y. Sakai and K. Hukushima, J. Phys. Soc. Jpn. {\bf 82}, 064003 (2013).

\bibitem{IO1}
A. Ichiki and M. Ohzeki, Phys. Rev. E {\bf 88}, 020101(R) (2013). 

\bibitem{IO2}
A. Ichiki and M. Ohzeki, Phys. Rev. E {\bf 91}, 062105 (2015).

\bibitem{OI}
M. Ohzeki and A. Ichiki, Phys. Rev. E {\bf 92}, 012105 (2015).

\bibitem{OI2}
M. Ohzeki and A. Ichiki, J. Phys.: Conf. Ser., {\bf 638}, 012003 (2015).

\bibitem{Liu}
J. S. Liu, Stat. Comput. {\bf 6}, 113 (1996); Biometrika {\bf 83}, 681 (1996).

\bibitem{PRHH}
L. Pollet, S. M. A. Rombouts, K. Van Houcke, and K. Heyde, 
Phys. Rev. E {\bf 70}, 056705 (2004).

\bibitem{SO}
Y. Sughiyama, and M. Ohzeki, 
Interdis. Info. Sci. {\bf 19}, 93 (2013).

\bibitem{CHKO}
A. Carlini, A. Hosoya, T. Koike, and Y. Okudaira, 
Phys. Rev. Lett. {\bf 96}, 060503 (2006).

\bibitem{CHKO2}
A. Carlini, A. Hosoya, T. Koike, and Y. Okudaira, 
J. Phys. A: Math. Theor. {\bf 41}, 045303 (2008).

\bibitem{DR1}
M. Demirplak and S. A. Rice, J. Phys. Chem. A {\bf 107}, 9937 (2003).

\bibitem{DR2}
M. Demirplak and S. A. Rice, J. Phys. Chem. B {\bf 109}, 6838 (2005).

\bibitem{Berry}
M. V. Berry, J. Phys. A: Math. Theor. {\bf 42}, 365303 (2009).

\bibitem{CRSCGM}
X. Chen, A. Ruschhaupt, S. Schmidt, A. del Campo, D. Gu\'ery-Odelin,  
and J. G. Muga, Phys. Rev. Lett. {\bf 104}, 063002 (2010).

\bibitem{STA}
E. Torrontegui, S. Ib\'a\~nez, S. Mart\'inez-Garaot, M. Modugno, 
A. del Campo, D. Gu\'ery-Odelin, A. Ruschhaupt, X. Chen, and J. G. Muga,
Adv. At. Mol. Opt. Phys. {\bf 62}, 117 (2013).

\bibitem{Takahashi}
K. Takahashi, J. Phys. A: Math. Theor. {\bf 46}, 315304 (2013).

\bibitem{LR}
H. R. Lewis and W. B. Riesenfeld, J. Math. Phys. {\bf 10}, 1458 (1969).

\bibitem{KGV}
S. Kirkpatrick, C. D. Galatt, and M. P. Vecchi, Science {\bf 220}, 671 (1983).

\bibitem{NDGD}
M. A. Nielsen, M. R. Dowling, M. Gu, and A. C. Doherty, 
Science {\bf 311}, 1133 (2006).

\bibitem{WH}
M. Welling and G. E. Hinton, Artificial Neural Networks-ICANN 2002, 
Lecture Notes in Computer Science {\bf 2415}, 351 (2002).

\bibitem{SBD}
J. Sohl-Dickstein, P. B. Battaglino, and M. R. DeWeese, 
Phys. Rev. Lett. {\bf 107}, 220601 (2011).

\bibitem{Ohzeki}
M. Ohzeki, arXiv:1511.06036 [stat.ML].

\end{thebibliography}
\end{document}